\documentclass[
    ,final            
    ,sort&compress
  ]
  {aipproc}

\usepackage{graphicx,amsmath}
\layoutstyle{6x9}

\newcommand{\tr}{\mathrm{Tr}}

\begin{document}

\title{MaxEnt assisted MaxLik tomography}

\author{J. \v{R}eh\'{a}\v{c}ek}{address={Department of Optics, 
Palacky University,
17. listopadu 50, 772 00 Olomouc, Czech Republic}}

\author{Z. Hradil}{
  address={Department of Optics, Palacky University,
17. listopadu 50, 772 00 Olomouc, Czech Republic}
}

\begin{abstract}
Maximum likelihood estimation is a valuable tool
often applied to inverse problems in quantum theory.
Estimation from small data sets can, however,
have non unique solutions. 
We discuss this problem and propose to use
Jaynes maximum entropy principle to single
out the most unbiased maximum-likelihood guess.
\end{abstract}

\maketitle


\section{Introduction}

The role of the variational principles in science can hardly be
overemphasized. Maximization or minimization of the appropriate
functionals provides elegant solutions of rather complicated problems
and  contributes to the deeper philosophical understanding of the
laws of Nature. Minimization of the optical path-Fermat principle
or minimization of the action-Hamilton principle are  two
particular examples of such a treatment in optics and classical
mechanics, respectively. In the thermodynamics and statistical
physics an appropriate measure which deserves to be maximized was
introduced by Boltzmann as entropy $ S = - k_b \log \Gamma$, where
$\Gamma$ denotes the volume in the phase space or the number of
distinguishable states. The role of entropy as uncertainty measure
in communication and information theory was recognized by Shannon.
His definition $S = - \sum_n p_n \log p_n $ is unique in  the
sense  that  fulfills reasonable demands put on the information
measure associated with a probability distribution $p_n$.
Particularly, the  uncertainty is maximized when all the outcomes
are equally likely-- the uniform distribution contains the
largest amount of uncertainty. Its implications for physical and
technical practice were noticed by Jaynes \cite{Jaynes}, who proposed a
variational method known as principle of Maximum Entropy (MaxEnt).
According to this rule one should select such a probability
distribution which fulfills given constraints and simultaneously
maximizes Shannon entropy. This gives the most unbiased
solution of the problem consistent with the given observations. On
the philosophical level this corresponds to the celebrated
Laplace's Principle of Insufficient Reasoning. It states that if
there is no reason to prefer among several possibilities, than the
best strategy is to consider them as equally likely and  pick up
the average. This strategy appeared to be extremely useful in many
applications covering the fields of statistical inference,
communication problems or pattern recognition \cite{frieden}.

But entropy is not the only important functional in probability theory 
The entropic measure known as Kullback-Leibler divergence 
\cite{Kullback} or relative entropy
$E(\{p_i\}|\{q_i\})  = \sum_i p_i \log (p_i/q_i) $
bears striking resemblance to the Shannon entropy, however it
posses a different interpretation. It quantifies the {\em distance} in
the statistical sense between two different distributions $p_i$ and $q_i$.
Provided that one party ($p_i$ in our notation) are the sampled
relative frequencies, the principle of minimum relative entropy
coincides with the maximum likelihood  (MaxLik) estimation
problem \cite{fisher,Helstrom}. 
Similarly to the previous case of MaxEnt principle MaxLik
is not a rule that requires justification - it does not need to be
proved. At present there are many examples of successful
application of this  estimation technique for solving inverse
problems, or recently, for quantification of such a fragile effect
as entanglement.

Though both the celebrated principles, MaxEnt and MaxLik, rely on
the notion of entropy, their usage and interpretation differ
substantially. Whereas the former one provides the most
conservative guess still consistent with the data, the later one
is the most optimistic one fitting the given data in the best
possible way \cite{frieden,fisher}.   
However, both the methods are suffering by certain
drawbacks: MaxLik is capable of dealing with counted noisy data in
realistic experiments but  its interpretation usually requires
a certain cut-off in the parameter space. Otherwise the solution may
appear us under-determined. On the other hand, the MaxEnt principle
removes this ambiguity by selecting the most unbiased solution,
however realistic data may appear as inconsistent due to the
fluctuations, and cannot be straightforwardly used as constraints.
The purpose of this contribution is to unify both these
concepts into a single estimation procedure capable of handling any
data, and to provide  the most likely and most unbiased solution
without any cut-offs.

\section{Maximum-likelihood quantum-state reconstruction}

To address the problem of  quantum state reconstruction
\cite{VR,St,SBRF93,Ulf,Welsch,buzek,banaszek}
let us
consider a generic quantum measurement. The formulation will be
developed for the case of  finite dimensional quantum systems. The
reader can think of a spin 1/2 system for the
sake of simplicity. 

Assume  that we are given a finite number $N$ of identical samples
of the system, each in the same but unknown quantum state described
by the density operator $\rho$. Given those systems our task is to
identify the unknown {\em true} state $\rho$ as accurately as possible
from the results of measurements performed on them. 

On most general level any set of measurements can be
represented by a Probability Operator Valued Measure (POVM),
$\{\Pi_j\},\, j=1\ldots M$. Its elements are semi-positive definite 
operators that sum up to unity operator, $\Pi_j\ge 0,\,\forall j$,
$\sum_j \Pi_j=\hat 1$. The last requirement is
simply the consequence of the conservation of probability:
The measured particle is always detected in one of the $M$ 
output channels, no particles are lost.

 Let us assume, for concreteness, that 
$N$ particles prepared in the same state have been
 observed in $M$ different output channels of the measurement apparatus.
 For spin 1/2 particles those channels could be for instance
 the six output channels of a Stern-Gerlach apparatus
 subsequently oriented along $x$, $y$, and $z$ directions.
 
Provided that each particular output
\begin{equation}  \label{projekce}
\Pi_j, \qquad j=1,\dots,M
\end{equation}
has been registered $n_j$ times, $\sum_j n_j = N,$ the relative
frequencies are given as $f_j = n_j/N.$ Using this data, the
true state $\rho$ is to be inferred. 
The probabilities of occurrences of various outcomes are
predicted by quantum mechanics as
\begin{equation} \label{teorie}
p_j=\tr\rho\Pi_j, \quad j=1\ldots M
\end{equation}

If the probabilities $p_j$ of getting a sufficient number of
different outcomes $\Pi_j$ were known, it would be possible
to determine the true state $\rho$  directly by inverting the
linear relation (\ref{teorie}). This is the philosophy behind the
``standard'' quantum tomographic techniques \cite{VR,Ulf}. For example, in the
rather trivial case of a spin one half particle, the probabilities of
getting three linearly independent projectors determine the
unknown state uniquely. Here, however, a serious problem arises.
Since  only a finite number of systems can be investigated, there
is no way how to find out those probabilities. The only data one
has at his or her disposal are the relative frequencies $f_j$, which sample
the {\em principally} unknowable probabilities $p_j$. It is
obvious that for a small number of runs, the true probabilities $p_j$
and the corresponding detected frequencies $f_j$ may differ
substantially. As a result of this, the modified realistic problem,
\begin{equation}    \label{problem}
f_j=\tr\Pi_j\rho
\end{equation}
has generally no solution on the space of semi-positive definite
hermitian operators describing physical states. This linear equation
for the unknown density matrix may be solved for example by means
of pattern functions, see e.g. \cite{Ulf,Welsch}, what could be
considered as a typical example of the standard approach
suffering from the above mentioned drawbacks.

Having measurements done and their results registered,
the experimenter's  knowledge about the measured system is increased.
Since quantum theory is probabilistic, it has little sense to ask the
question: "What quantum state is determined by that measurement?"
More appropriate question is \cite{Bayes-class,Bayes-quant,zdenek_fund,
zdenek_nelin}: "What quantum states seem to be most
likely for that measurement?"

Quantum theory predicts
the probabilities of individual detections, see Eq.~(\ref{teorie}). 
From them one can construct the total joint probability
of registering data $\{n_j\}$. 
We assume that the input system (particle) is
always detected in one of $M$ output channels,
and this is repeated $N$ times. 
Subsequently, the overall statistics of the experiment
is multinomial, 
\begin{equation} \label{MaxLik}
{\cal L}(\rho)=\frac{N!}{\prod_i n_i!} \prod_j 
\bigl[\tr(\rho\Pi_j)\bigr]^{n_j},
\end{equation}
where  $n_j = N f_j$  denotes the rate of registering a
particular outcome $j$. In the following we will omit the multinomial 
factor from expression~\eqref{MaxLik} as it has no influence on the results.
Physically, the quantum state reconstruction corresponds to a 
synthesis of  various measurements
done under different experimental conditions,  performed on the
ensemble of identically prepared systems. For example, the
measurement might be subsequent recording of an unknown spin of the
neutron (polarization of the photon)  using different settings of
the Stern Gerlach apparatus, or the recording of the quadrature
operator of light in rotated frames in quantum homodyne tomography. The
likelihood functional ${\cal L}(\rho)$ quantifies the degree of
belief in the hypothesis that for a particular data set $\{ n_j\}$ the
system was prepared in the quantum state $\rho$. The MaxLik
estimation simply selects the state for which the likelihood attains
its  maximum value on the manifold  of density matrices.

To make the mathematics simpler we will
maximize the logarithm of the likelihood functional, 
\begin{equation} \label{loglik}
L\propto\sum_j f_j \log p_j,
\end{equation}
rather then  $\mathcal{L}$ itself.
Notice that $L$ is a convex functional,
\begin{equation} \label{convexity}
L[\alpha\rho_1+(1-\alpha)\rho_2]\ge\alpha L(\rho_1)
+(1-\alpha)L(\rho_2),
\end{equation}
defined on the convex set of semi-positive definite density matrices
$\rho$, $\rho\ge 0$, $\tr\rho=1$.
This ensures that there is a single global maximum or at most
a closed set of equally likely quantum states.

The direct application of the variational principle
to likelihood functional together with the convexity property yield
the necessary and sufficient condition for its maximum in the form
of a nonlinear operator equation \cite{Hradil1,Rehacek01},
\begin{equation} \label{extremal}
R\rho=\rho,
\end{equation}
where 
\begin{equation} \label{operr}
R=\sum_j\frac{f_j}{p_j}\Pi_j
\end{equation}
is a semi-positive definite operator.
In particular, $R$ is unity operator provided the maximum-likelihood
solution is strictly positive.

Let us now consider a tomographically incomplete measurement.
In such a case, the inverse problem might have multiple
solutions. This will happen, for example, when the
set of normalized Hermitian operators $\sigma$ 
satisfying the constraints $p_j(\sigma)=f_j,\, \forall j$
has a nonempty intersection with the set of density matrices.
As will be illustrated below, two equally-likely 
solutions of an under-determined inverse problem can be very different.
It is then a question which maximum-likely state should be
picked up as the estimate of the true state.

We propose to use Jaynes MaxEnt principle to resolve
this ambiguity. Information content of the set of MaxLik solutions
can be quantified according to their entropy. A natural choice
is then to select the state maximizing the entropy, which
is the least biased state with respect to missing measurements.

Let us assume that there are two different density operators
$\rho_1$ and $\rho_2$ maximizing the likelihood functional.
The two operators satisfy the extremal equations \eqref{extremal},
\begin{equation} \label{multiple}
\begin{split}
R(\rho_1)\rho_1&=\rho_1,\\
R(\rho_2)\rho_2&=\rho_2.
\end{split}
\end{equation}

The interpretation of the operator $R$ is the following:
Denoting $f(\rho,\sigma,\alpha)\equiv L[(1-\alpha)\rho+\alpha\sigma]$
the likelihood  of  a convex combination of states $\rho$ and $\sigma$,
and calculating its path derivative at $\rho$,
\begin{equation} \label{deriv}
\begin{split}
\frac{\partial}{\partial\alpha}f(\rho,\sigma,\alpha)&=
\lim\limits_{\alpha\rightarrow 0}
\frac{f(\rho,\sigma,\alpha)-f(\rho,\sigma,0)}{\alpha}\\
&=\tr[R(\rho)\sigma]-1,
\end{split}
\end{equation}
we see that this derivative is given by the expectation value
of $R(\rho)$ taken with $\sigma$.
Expectation values of operator $R(\rho)$ 
define hyperplanes perpendicular to the gradient of
the likelihood at $\rho$.

Since the likelihood cannot be increased by moving from $\rho_1$ toward
$\rho_2$ and \textit{vice versa} (both density matrices are 
maximum likely states) it follows that
\begin{equation} \label{condition}
\tr[R(\rho_1)\rho_2]=\tr[R(\rho_2)\rho_1]=1.
\end{equation}

Expressing the two conditions in terms of probabilities 
$p_{1j}$ and $p_{2j}$ generated by $\rho_1$ and $\rho_2$, respectively
we get
\begin{equation} \label{condition2}
\begin{split}
&\sum_j f_j \frac{p_{1j}}{p_{2j}}=1,\\
&\sum_j f_j \frac{p_{2j}}{p_{1j}}=1,
\end{split}
\end{equation}
which upon summing the left-hand sides yields condition
\begin{equation} \label{ineq}
\sum_jf_j\frac{p_{1j}^2+p_{2j}^2}{2p_{1j}p_{2j}}= 1.
\end{equation}
Now since $(p_{1j}^2+p_{2j}^2)/(2p_{1j}p_{2j})>1$
unless $p_{1j}=p_{2j}$ we obtain,
\begin{equation} \label{equiprobab}
\tr\rho_1\Pi_j=\tr\rho_2\Pi_j, \quad \forall j
\end{equation}
which means that the probabilities, and so the operators $R(\rho_1)$ and 
$R(\rho_2)$ are identical.
The two extremal equations therefore read,
\begin{equation} \label{equir}
\begin{split}
R\rho_1&=\rho_1,\\
R\rho_2&=\rho_2.
\end{split}
\end{equation}
Notice that both $\rho_1$ and $\rho_2$ commute with the common generator $R$.

\section{Maximization of entropy}

Having found a maximum of the likelihood functional,
we still do not know whether this solution is unique or not.
Provided a closed set of such states exists,
we would like to maximize the entropy functional over it.
In this way we will get the least biased maximum-likelihood guess.

The properties of the maximum-likelihood solutions discussed
above simplify this problem a lot, because we know
that all density matrices belonging to the maximum likely set
generate the same probabilities.

We will take those probabilities as constraints of the
new optimization problem: Maximize entropy,
\begin{equation}\label{maxentproblem}
E(\rho)=-\tr(\rho\ln\rho),
\end{equation}
subject to constraints
\begin{equation}\label{maxentconstraints}
\tr(\rho\Pi_j)=\tr(\rho_\mathrm{ML}\Pi_j), \quad j=0\ldots M,
\end{equation}
where $\rho_\mathrm{ML}$ is a maximum likely state and where 
we defined $\Pi_0=\hat 1$ to keep the normalization of the
estimated state.

Problem Eq.~\eqref{maxentproblem} and \eqref{maxentconstraints}
is known to have a unique solution \cite{Jaynes},
\begin{equation} \label{jaynes}
\rho_\mathrm{ME}=\exp\Bigl[\sum_j\lambda_j\Pi_j\Bigr], \quad j=0\dots M,
\end{equation}
where Lagrange multipliers $\lambda_j$ can be determined from
the constraints.

The proposed approach combines good features of maximum-likelihood
and maximum-entropy methods. From the set of density matrices
that are most consistent with the observed data in the sense
of maximum likelihood we select the least biased one.
At the same time the positivity, and thus also physical soundness,
of the result is guaranteed.

Let us remind the reader that a direct application of the maximum entropy
principle to raw data (i.e. right hand sides in constraints
Eq.~\eqref{maxentconstraints} replaced by $f_j$) is not 
possible, because the constraints often cannot be 
satisfied with any semi-positive density operators due to
the unavoidable presence of noise in the data.

For the rest of the paper let us restrict ourselves to
the most simple case of commuting measurements 
$[\Pi_j,\Pi_k]=0,\;\forall j,k$.
Such tomographic scheme would correspond to the measurement
of diagonal elements of the true density matrix.
Although this may seem as an oversimplification,
many inverse problems can be reduced to this form.
Let us mention the neutron absorption tomography,
or inefficient photo detection as two significant examples.

\section{Example}

We will illustrate the proposed reconstruction scheme 
on a simple example of commuting measurements.
Denoting $\{|i\rangle\langle i|\}$ the common eigenbasis
of POVM elements $\{\Pi_j\}$ we have that
\begin{equation} \label{equations}
\sum_j \lambda_j\Pi_j=\sum_i r_i|i\rangle\langle i|.
\end{equation}
\begin{figure}
\centerline{
\includegraphics[width=0.6\columnwidth]{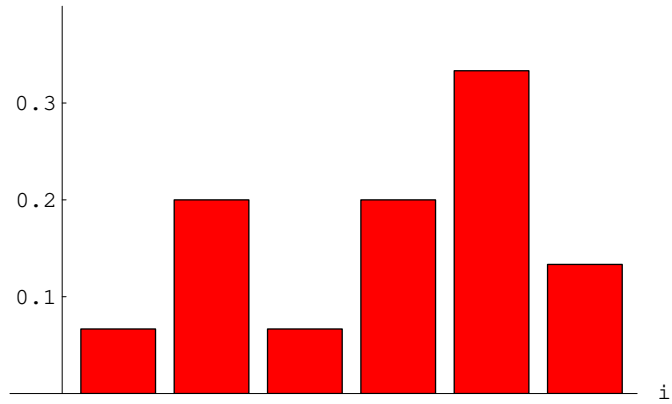}
}
\caption{Diagonal representation of a particular true state.\label{fig:true}}
\end{figure}
The maximum entropy solution Eq.~\eqref{jaynes} will assume
a diagonal form in this basis, its eigenvalues being, 
\begin{equation}
\langle i|\rho_\mathrm{ME}|i\rangle=\exp\bigl[\sum_j\lambda_j
\langle i|\Pi_j|i\rangle\bigr].
\end{equation}
Denoting $\rho_i=\langle i|\rho|i\rangle$ and 
$c_{ij}=\langle i|\Pi_j|i\rangle$ we finally get
a nonlinear system of equations,
\begin{equation}
\sum_i e^{\sum_{j'}\lambda_{j'}c_{ij'}}c_{ij}=
\sum_i c_{ij}\rho_{\mathrm{ML},i}\quad,
\end{equation}
that is to be solved for the unknown $M+1$ Lagrange multipliers
$\lambda_j$. 

\begin{figure}
\centerline{
\includegraphics[width=0.6\columnwidth]{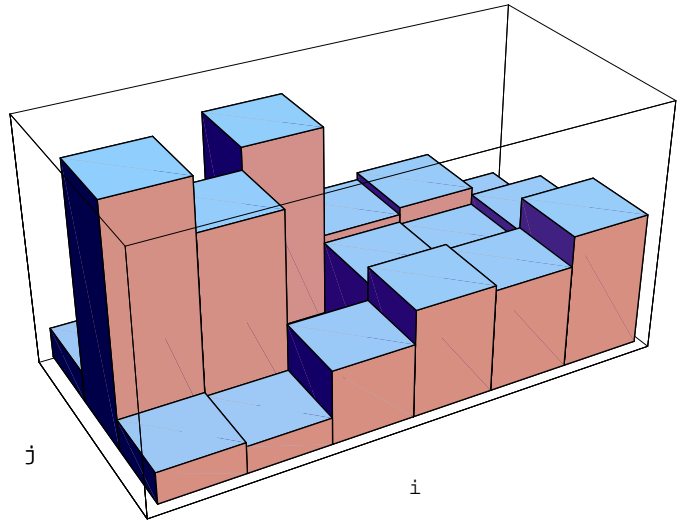}
}
\caption{Randomly generated matrix $c_{ij}$ parameterizing 
a three element POVM.\label{fig:kernel}}
\end{figure}

A particular true six-dimensional vector $\rho_{\mathrm{true},i}$
is shown in Fig.~\ref{fig:true}.
In a simulated experiment this state has been subject to
randomly generated three element $POVM$; its elements $c_{ij}$
in the common diagonalizing basis are shown in Fig.~\ref{fig:kernel}.

\begin{figure}
\centerline{
\includegraphics[width=0.6\columnwidth]{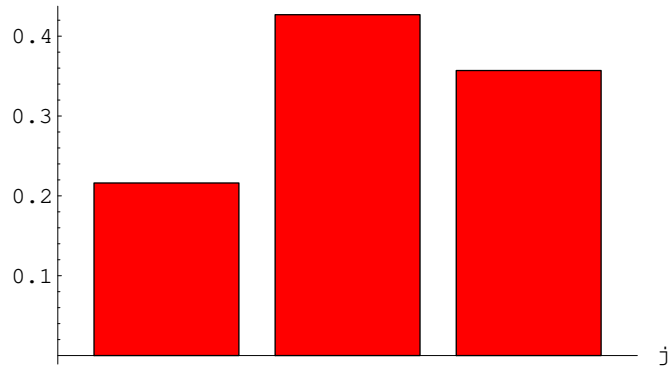}
}
\caption{Relative frequencies of the outcomes
of a thought tomographic measurement. For the true state and
POVM elements, see Figs.~\ref{fig:true} and \ref{fig:kernel}.
\label{fig:counts}}
\end{figure}

The probabilities of observing results $j=1,2,3$ 
are as follows: $p_j=\sum_i c_{ij}\rho_{\mathrm{true},i}$.
They are shown in Fig.~\ref{fig:counts} for our particular
choice of $\rho_\mathrm{true}$ and $c_{ij}$.
Taking the probabilities as the input data, we solved
the maximum-likelihood extremal equation iteratively
starting from three different strictly positive density matrices.
It is worth noticing that a quantum state reconstruction from compatible 
observations is a linear and positive problem.
In this case the operator equation \eqref{extremal}
reduces to a simple diagonal form which is suitable to iterative solving.
This algorithm is sometimes called  
the expectation maximization algorithm in statistical literature
and is known to converge monotonically from any strictly positive initial
point \cite{dempster,vardi} .

\begin{figure}
\centerline{
\includegraphics[width=0.6\columnwidth]{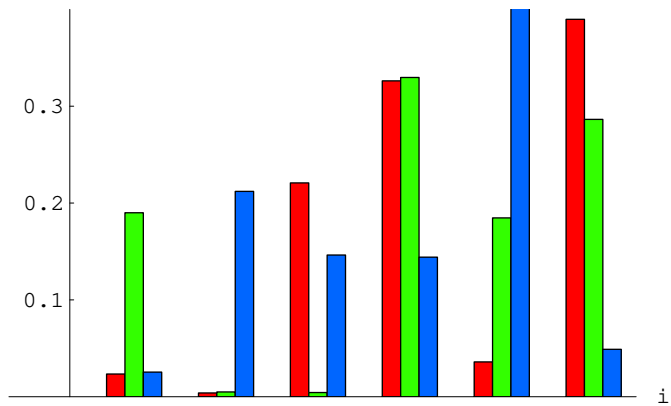}
}
\caption{Three particular maximum-likelihood estimates
based on the result of the though experiment
described in the text. \label{fig:maxlik}}
\end{figure}

As we can see, the three maximum-likely estimates
represent very different system configurations.
The simple POVM used was too rough to resolve those
differences, and as a consequence, all the estimated states yield exactly the
same probabilities of the three possible outcomes of the measurement.

\begin{figure}
\centerline{
\includegraphics[width=0.6\columnwidth]{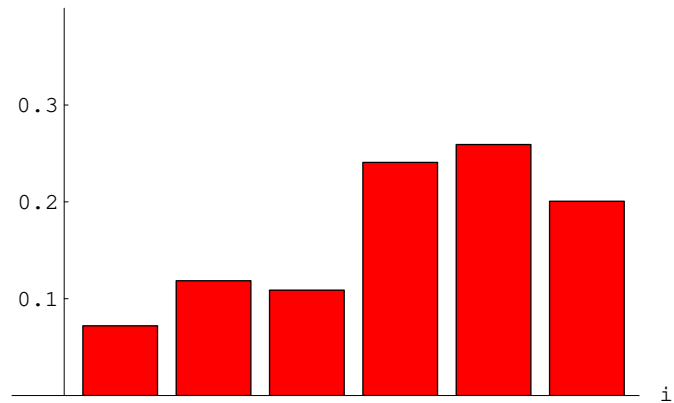}
}
\caption{The result of entropy maximization over the set
of maximum-likelihood estimates (three of which are shown
in Fig.~\ref{fig:maxlik}.)\label{fig:maxent}}
\end{figure}

In the next step, those probabilities were used as
constraints for the entropy maximization, as we have discussed
in the previous section. As a result, a 
unique state was selected out of the set of 
maximum-likely states. The result is shown in Fig.~\ref{fig:maxent}.
Notice that this state is a good approximation to the original
state of Fig.~\ref{fig:true}. Even though the two
are smoothed out a bit, they can be clearly recognized
in the reconstruction.

\section{Conclusion}

We have demonstrated the utility of the maximum-entropy
principle for tomographically incomplete quantum state
reconstruction schemes. Although the entropic principles cannot
be directly applied to noisy experimental data 
due to the positivity of quantum states,
they can be used to remove the ambiguity of maximum likelihood estimation.
The proposed method could find applications in quantum homodyne
detection and other related infinite-dimensional problems 
suffering from the lack of experimental data.

\begin{theacknowledgments}
This work was supported by the projects LN00A015 and CEZ:J14/98 
of the Czech Ministry of Education and Czech-Italian project 
``Decoherence and quantum measurement.''
\end{theacknowledgments}


\end{document}